\documentclass[12pt]{amsart}
\usepackage{amssymb}
\usepackage{amsmath}
\usepackage{graphicx}
\usepackage{soul}
\usepackage{color}

\begin{document}
\title{From chaos to order through mixing}
\author{D. DMITRISHIN,  I.M. SKRINNIK and A.STOKOLOS }
 
\begin{abstract}
     In this article we consider the possibility of controlling the dynamics of nonlinear discrete systems. A new method of control is by mixing states of the system (or the functions of these states) calculated on previous steps. This approach allows us to locally stabilize a priori unknown cycles of a given length. As a special case, we have a cycle stabilization  using nonlinear feedback. Several examples are considered. The earlier version of this article is published in \cite{0} \\
 
       Keywords: non-linear discrete systems, chaos control, mixing of system states.
\end{abstract}
\maketitle
\section{Introduction}
 
%%%%%%%%%%%
The notion of chaos control was introduced in \cite{1}, chaos control being the  concept of stabilization of unstable periodic orbits of non-linear systems by small controlling impacts that however totally change the nature of the motion. The aim of such control is either synchronization of a chaotic motion or chaotization of a regular motion. Later on several other methods of chaos control were suggested. The most popular turned out to be the DFC method by Pyragas \cite{6}. The idea of the method is introduction of linear feedback with delay of the size of period. The Pyragas method got its popularity in many respects due to its constructive simplicity. Further, the method was developed and generalized in many papers, c.f. \cite{2},\cite{3},\cite{4}. There are known other various approaches to control, e.g. \cite{5}-\cite{13}. However, each of them has significant deficiency \cite{14a}. Thus the problem of development of new methods of chaos control still remains.\\
 
In \cite{14,15}, to stabilize periodic orbits the application of non-linear feedback with several delays of size proportional to the period were suggested. Developing this methodology, in this paper is suggested a new method of solving the problem of optimal stabilization of cycles in families of discrete autonomous systems. This method involves  mixing the previous stages of the system or functions of these stages. In particular cases the method of mixing coincides with control by means of delay feedback \cite{2,14,15}. \\
 
{\it The goal of this paper} is a construction of algorithms of chaos suppression in non-linear discrete systems, based on a newly considered class of problems of physical principles. \\
 
{\it The issue} is in the choice of structure and parameters of the control system, that locally stabilize a priori unknown cycles of given length.
 
\section{Settings and preliminary results}
 
Let us consider the vector nonlinear discrete system that without control has a form
\begin{equation}\label{1}
x_{n+1} =f\left(x_{n} \right),\, \, x_{n} \in\mathbb R^{H} ,\, \, n=1,\, \, 2,\, \, \ldots \, \, ,
\end{equation}
where $f(x)$ is a differentiable vector function of appropriate dimension.\\
 
It is assumed that the system \eqref{1} has invariant convex sets $A,$ i.e. if $\xi\in A$ then $f(\xi)\in A.$ It is also assumed that this system has one or several unstable $T-$ cycles $\left(\eta_{1},\, \, \ldots \, , \eta_{T}\right)$, where all vectors $\eta_{1},\, \, \ldots \, ,  \eta_{T}$ are pairwise distinct and belongs to the invariant set $A,$ i.e.
$\eta_{j+1}=f(\eta_{j}), j=1,\ldots,T-1, \eta_{1}=f(\eta_{T})$.\\
 
The multipliers considered  for the unstable cycles are determined as eigenvalues of a product of Jacobi's matrices $\prod_{j=1}^T f'(\eta_j)$ with $H$ dimension.\\
 
  It is required to construct the system
\begin{equation}\label{2}
x_{n+1}=\sum_{j=1}^M\gamma_j f(\sum_{i=1}^N \alpha_{ij}x_{n-iT+T}),
  \end{equation}
  where
$$
\gamma_j\geqslant 0, \alpha_{ij}\geqslant 0,
\sum_{j=1}^M \gamma_j = 1, \sum_{i=1}^N \alpha_{ij}=1, j=1,\ldots, M,\quad M\in \mathbb N.
$$
The goal of control is to make locally stable all (or at least some) $T-$ cycles of the system \eqref{2}. It is important that the convex set $A$ still invariant for the system
\eqref{2} as well , i.e. if vectors $\xi_0,\xi_T,\xi_{2T},\ldots,\xi_{(N-1)T}$ are contained in set A, then the vector $\sum_{j=1}^M\gamma_j f(\sum_{i=1}^N \alpha_{ij}\xi_{(N-i)T})$ is contained in the set $A$ too. This follows from the definition of the convex combination of the vectors. On a top of that the system \eqref{2} has same $T$-cycles that is in the system \eqref{1}.\\
 
A physical meaning of the control is in the following:  the expression $\sum_{i=1}^N \alpha_{ij}x_{n-iT+T}$ can be interpreted as inside mixing of system states $x_n,x_{n-T},\ldots,x_{n-(N-1)T}$ while the expression $\sum_{j=1}^M\gamma_j f(x_{n-jT+T})$ can be interpreted as outside mixing. Thus,\\ $\sum_{j=1}^M\gamma_j f(\sum_{i=1}^N \alpha_{ij}x_{n-iT+T})$ is a combined mixing. Note that control with outside mixing is equivalent to the delayed feedback control \cite{15}.\\

Let us  apply the following linearization to construct  Jacobi matrices for the system \eqref{2} and its characteristic equation. This scheme was justified in \cite{16,17}.  \\
 
It is clear that
\begin{equation}
  \begin{cases}
x_{n+k}=\sum_{j=1}^{M}\gamma_j f(\sum_{i=1}^N \alpha_{ij}x_{n+k-1-iT+T})\\
k=1,\ldots,T
\end{cases}
\end{equation}
Solution of  the system (3) can be written  in the form
\begin{equation}
\begin{split}
&x_{Tm}=\eta_1+u_m^1 \\
&\dots \dots \dots\dots\dots \\
&x_{Tm+T-1}=\eta_T+u_m^T
\end{split}
\end{equation}
Let us substitute the solution (4)  in (3) assuming that values $u_m^1,\ldots,u_m^T$ are so small in the neighborhood of a cycle that the assumptions of the theorem by first approximation is valid \cite{1}.\\

  Let $n=Tm$. Then
$$
x_{n+1}=x_{Tm+1}=\eta_2+u_m^2,x_{n+2}=x_{Tm+2}=\eta_3 +u_m^3,\ldots, x_{n+T}=x_{T(m+1)}=\eta_1+u_{m+1}^1.
$$
  Separating the linear part and given that $ \eta_1=f(\eta_2),\ldots,\eta_T=f(\eta_1)$, we obtain
   \begin{equation}
   \begin{split}
   & u_m^2=f'(\eta_1)(a_1u_m^1+\ldots+a_Nu_{m-N+1}^1)\\
   & u_m^3=f'(\eta_2)(a_1u_m^2+\ldots+a_Nu_{m-N+1}^2)\\
   & \dots \dots \dots\dots\dots\dots\dots\dots\dots\dots\dots\dots \,\,\,\,\,\,\,\,\,\,\,\,,\\
   & u_m^T=f'(\eta_{T-1})(a_1u_m^{T-1}+\ldots+a_Nu_{m-N+1}^{T-1})\\
   & u_{m+1}^1=f'(\eta_T)(a_1u_m^T+\ldots+a_Nu_{m-N+1}^T)\\
   \end{split}
   \end{equation}
   where
    \begin{equation}
    a_i=\sum_{j=1}^{M}\alpha_{ij}\gamma_j, i=1,\ldots,N.
    \end{equation}
The system (5) is linear therefore the solutions are exponents, i.e.
$$
\left(\!\begin{array}{c}u_m^1\\\dots\\u_m^T\end{array}\!\right)=\left(\!\begin{array}{c}c_1\\\dots\\c_T\end{array}\!\right)\lambda^m\,\,\,,
$$
that after substitution in (5) give us a system
  \begin{equation}
  \begin{split}
&\left(\!\!\!\!\begin{array}{cccccc}-f'(\eta_1)\cdot p(\lambda^{-1})&I&0&\dots&0&0\\0&-f'(\eta_2)\cdot p(\lambda^{-1})&I&\dots&0&0\\\dots&\dots&\dots&\dots&\dots&\dots\\0&0&0&\dots&-f'(\eta_{T-1})\cdot p(\lambda^{-1})&I\\\lambda I&0&0&\dots&0&-f'(\eta_T)\cdot p(\lambda^{-1})\end{array}\!\!\!\!\right)\times\\
&\times\left(\!\begin{array}{c}c_1\\c_2\\\dots\\c_{T-1}\\c_T\end{array}\!\right)=\left(\!\begin{array}{c}0\\0\\\dots\\0\\0\end{array}\!\right)
\end{split}
  \end{equation}
where $p(\lambda^{-1})=a_1+a_2\lambda^{-1}+\ldots+a_N\lambda^{-N+1}$, $I$ is identity matrix and $0$ is a null matrix of dimensions $H$.\\
 
The standard methods of analyzing $T$-cycle stability is verification that the location of all zeros of the determinant of system (7) is in the central unit disc of the complex plane $\mathbb D=\{z:|z|<1\} $. In our case the determinant of system (7) is equal to
$$
\begin{aligned}
&{\rm det}((-1)^{T-1}\lambda I+\prod_{j=1}^T (-f'(\eta_j)p(\lambda^{-1})))={\rm det}(\lambda I-(p(\lambda^{-1}))^T\prod_{j=1}^T f'(\eta_j))=\\&=\prod_{j=1}^H(\lambda-\mu_j(p(\lambda^{-1}))^T),
\end{aligned}
$$
where $\mu_1,\ldots,\mu_H$ are eigenvalues of matrix $\prod_{j=1}^Tf'(\eta_j)$.
 
\section{The problem of optimal mixing}
 
The $T$-cycle stability condition of system (2) is a requirement that all roots of the equation
\begin{equation}
\prod_{j=1}^H(\lambda^{1+(N-1)T}-\mu_j(a_1\lambda^{N-1}+\ldots+a_N)^T)=0
\end{equation}
are in the central unit circle. Corresponding, the problem of chaos control in system (1) by mixing values of system's state and function of this values in previous moments of time is formulated in the follwing way: {\it for given cycle length $T$ and given set of multipliers localization define the coefficients of inside mixing $\alpha_{ij},i=1,\ldots,N,j=1,\ldots,M $ , and outside mixing $\gamma_{j},j=1,\ldots,M$, such that cycle of length $T$ will be locally asymptotically stable; the magnitude of using prehistory should be minimum possible.}
 
In other words, in equation (8) it is necessary to find the coefficients $a_1,\ldots,a_N$ such that all roots of this equation lie in the central unit circle, and the value $N$ should be the minimum possible.\\
 
Clearly, the solution of problem depends on the localization of multipliers  $\{\mu_1,\ldots,\mu_H\}$.\\
 
We will consider two possibilities: either all multipliers are real
$$\{\mu_1,\ldots,\mu_N\}\subset\{\mu\in \mathbb R:\mu\in(-\mu^*,1)\},\; \mu^*>1,$$
or  are in the left half-plane
$$\{\mu_1,\ldots,\mu_H\}\subset\{\mu\in \mathbb C:|\mu+R|<R\},\; R>1/2.$$
For each of these cases the algorithm of finding minimum $N$ and optimal coefficients $\{a_1,\ldots,a_N\}$ consists of the following steps \cite{18}:
 
\begin{itemize}
                \item [a)] compute nodes:
                $$
                \psi_j=\frac{{\pi(\sigma+T(2j-1))}}{\sigma+(N-1)T}, j=1,2,\ldots,\frac{N-2}{2},\;\mbox{$N$ even;\; $j=1,2,\ldots,\frac{N-1}{2}$, $N$ odd};
                $$
                In the case $\{\mu_1,\ldots,\mu_H\}\subset\{\mu \in\mathbb R:\mu\in(-\mu^*,1)\}$ we let $\sigma=2$, while in the case $\{\mu_1,\ldots,\mu_H\}\subset\{\mu\in \mathbb C:|\mu+R|<R\}$  we let $\sigma=1$;
                \item[b)] construct auxiliary polynomials
                $$
                \eta_N(z)=z(z+1)\prod_{j=1}^{\frac{N-2}{2}}(z-e^{i\psi_j})(z-e^{-i\psi_j}), \;\mbox{$N$ even},
                $$
                $$
                \eta_N(z)=z\prod_{j=1}^{\frac{N-2}{2}}(z-e^{i\psi_j})(z-e^{-i\psi_j}),\;\mbox{ $N$ odd;}
                $$
                \item[c)] compute coefficients of auxiliary polynomials
                $$
                \eta_N(z)=\sum_{j=1}^{N}c_jz^j;
                $$
                \item[d)] construct optimal coefficients
                $$
                a_j=\frac{(1-\frac{1+(j-1)T}{2+(N-1)T})c_j}{\sum_{j=1}^{N}(1-\frac{1+(j-1)T}{2+(N-1)T})c_j},\; j=1,\ldots,N;
                $$
                \item[e)] in case $\{\mu_1,\ldots,\mu_N\}\subset\{\mu\in \mathbb R:\mu\in(-\mu^*,1)\}$ compute values
                $$
                I_N^{(T)}=\left[\frac{T}{2+(N-1)T}\prod_{k=1}^{\frac{N-2}{2}}\cot^2\frac{\pi(2+T(2k-1))}{2(2+(N-1)T)}\right]^{T},\;\mbox{$N$ even}
                $$
                $$
                I_N^{(T)}=\left[\prod_{k=1}^{\frac{N-2}{2}}\cot^2\frac{\pi(2+T(2k-1))}{2(2+(N-1)T)}\right]^{T},\; \mbox{$N$ odd};
                $$
                The optimal value of $N$ is computed as minimal positive integer that satisfies the inequality $\mu^*\cdot I_N^{(T)}<1$;
                \item[f)] in case $\{\mu_1,\ldots,\mu_H\}\subset\{\mu\in \mathbb C:|\mu+R|<R\}$ compute values
                $$
                I_N^{(T)}=\left[\frac{T}{1+(N-1)T}\prod_{k=1}^{\frac{N-2}{2}}\cot^2\frac{\pi(1+T(2k-1))}{2(1+(N-1)T)}\right]^{T},\;\mbox{$N$ even},
                $$
                $$ 
                I_N^{(T)}=\left[\prod_{k=1}^{\frac{N-2}{2}}\cot^2\frac{\pi(1+T(2k-1))}{2(1+(N-1)T)}\right]^{T},\;\mbox{$N$ odd;}
                $$
                The optimal value of $N$ is computed as minimal positive integer that satisfies the inequality $R\cdot 2I_N^{(T)}<1$.
\end{itemize}
 
If the optimal coefficients are found then the mixing coefficients can be found from the system (6), which can be conveniently written as
\begin{equation}
\left(
\begin{array}{cccc}
\alpha_{11}&\alpha_{12}&...&\alpha_{1M}\\
\alpha_{21}&\alpha_{22}&...&\alpha_{2M}\\
...&...&...&...\\
\alpha_{N1}&\alpha_{N2}&...&\alpha_{NM}\\
\end{array}
\right)\left(\begin{array}{c}
\gamma_1\\
\gamma_2\\
...\\
\gamma_M\\
\end{array}\right)=\left(\begin{array}{c}
a_1\\
a_2\\
...\\
a_N\\
\end{array}\right)
\end{equation}
If we add normalization condition than the system takes the form
$$\left(
\begin{array}{cccc}
\alpha_{11}&\alpha_{12}&...&\alpha_{1M}\\
...&...&...&...\\
\alpha_{N-1,1}&\alpha_{N-1,2}&...&\alpha_{N-1,M}\\
1&1&1&1\\
\end{array}
\right)\left(\begin{array}{c}
\gamma_1\\
\gamma_2\\
...\\
\gamma_M\\
\end{array}\right)=\left(\begin{array}{c}
a_1\\
...\\
a_{N-1}\\
1\\
\end{array}\right)$$
Let us note that the optimal coefficients are determined uniquely while mixing coefficients are not necessary unique.
 
\section{Examples of systems with mixing}
Let us consider a few examples.\\
 
   Let $M=N$ and $\alpha_{ij}=\delta_{ij}$, where $\delta_{ij}$ - Kroneker's symbol. Then, from (9) we get $\gamma_j=a_j,\,\,\,\, j=1,\ldots,N$. The system (2) takes the form $$ x_{n+1}=\sum_{j=1}^{M}a_jf(x_{n-jT+T}).$$
The system is controlling by the outside mixing \cite{15}.\\
 
Let $M=1$, then $\gamma_1=1$, and $\alpha_{i1}=a_i$, $i=1,\ldots,N$. The system (2) takes the form $$x_{n+1}=f(\sum_{i=1}^{N}a_ix_{n-iT+T}).$$
The system is controlling by inside mixing.\\
 
Let $M=N+1$; $\alpha_{i1}=a_i$, $i=1,\ldots,N$; $\alpha_{i,i+1}=1$, $i=1,\ldots,N$; $\alpha_{ij}=0$, where $j\ne1$, $j\ne i+1$, $i=1,\ldots,N$, $j=1,\ldots,N+1$. Then the system (2) takes the form
$$ x_{n+1}=\gamma_1f(\sum_{i=1}^{N}a_ix_{n-iT+T})+
\sum_{j=1}^{N}\gamma_{j+1}f(x_{n-jT+T}).
$$
We determine the optimal values of coefficients of outside mixing $\gamma_1,\ldots, \gamma_{N+1}$.
In this case the system (9) takes the form
 
$$\left(\begin{array}{ccccc}
a_1&1&0&\ldots&0\\
a_2&0&1&\ldots&0\\
\ldots&\ldots&\ldots&\ldots&\ldots\\
a_N&0&\ldots&\ldots&1\\
\end{array}\right) \left(\begin{array}{c}
\gamma_1\\
\gamma_2\\
\ldots\\
\gamma_{N+1}\\\end{array}\right)=\left(\begin{array}{c}
a_1\\
a_2\\
\ldots\\
a_N\\\end{array}\right).$$
The general solution of this system is represented as the sum of the partial solution of the inhomogeneous system and the general solution of the homogeneous system
$$\left(\begin{array}{c}
\gamma_1\\
\gamma_2\\
\ldots\\
\gamma_{N+1}\\\end{array}\right)=\left(\begin{array}{c}
0\\
a_1\\
\ldots\\
a_{N}\\\end{array}\right)+c\left(\begin{array}{c}
1\\
-a_1\\
\ldots\\
-a_{N}\\\end{array}\right).
$$
Given that $\gamma_j\ge 0$, $j=1,\ldots,N+1$, we get $0\le c\le 1$. Finally the system (2) takes the form
\begin{equation}\label{sysc}
x_{n+1}=c\,f(\sum_{i=1}^{N}a_ix_{n-iT+T})+(1-c)\sum_{j=1}^{N}a_jf(x_{n-jT+T}),
\end{equation}
where $0\le c\le 1$. The system is controlled with help of combined mixing. When $c=0$ we do outside mixing, when $c=1$ we do inside mixing. The systems \eqref{sysc} appears e.g. in diffusion chaos theory, c.f. \cite{19}.
 
\section{Multiples cycles}
 
In this section we will focus on the possibility of controlling the chaos by stabilizing $lT$-cycles with mixing, determined by the dynamic system (2), $l=2,3,\ldots$.
 
Each point $T$-cycle is a fixed point of the $T$-iterated mapping $f$:
$$
x_{n+1}=f^{(T)}(x_n),\,\,\,\,f^{(T)}(x_n)=f(f^{(T-1)}(x_n)),\,\,\,\,f^{(0)}(x_n)=f(x_n).
$$
The problem of the cycle stability is reduced to the question of the fixed points stability of the mapping $f^{(T)}$, that generates the cycle of length $T$. Note, that the value of cycle multiplier does not depend on the choice of the fixed point that is included in the considered cycle, and that the fixed points of the mapping $f^{(T)}$ are also fixed points of the mapping $f^{(T_1)}$, if $T_1=lT$ with $l$ being an integer.
 
Thus, the problem of stabilization of $T$-cycles of system (1) can be reduced to the problem of stabilization of $T$-cycles for the system
$$
x_{n+1}=f^{(l)}(x_n),\,\,x_n\in \mathbb R^H,\,\,n=1,2,\ldots.
$$
Mixing will be organized as following.\\
 
The mixing of the first level: $$\sum_{i=1}^{N}\alpha_ix_{n-iT+T};$$
 
Then the mixing of the second level: $$\sum_{j=1}^{M}\beta_jf(\sum_{i=1}^{N}\alpha_{ij}x_{n-iT+T});$$
 
Then the mixing of the third level: $$\sum_{j_1=1}^{M_1}\gamma_{j_1}f(\sum_{j_2=1}^{M_2}\beta_{j_2}
f(\sum_{i=1}^{N}\alpha_{ij_2j_1}x_{n-iT+T})).$$
 
The mixing of  the higher levels is determined similarly.\\
 
Thus, the control system for the stabilization of cycle length $T$ is represented as the following
\begin{equation}
x_{n+1}=\sum_{j_1=1}^{M_1}\gamma_{j_1}f(\sum_{j_2=1}^{M_2}\beta_{j_2}
f(\sum_{i=1}^{N}\alpha_{ij_2j_1}x_{n-iT+T})).
\end{equation}
In the system (10), all coefficients have to satisfy the normalization conditions, thus define the convex combinations.\\
 
The mixing coefficients can be considered as components of tensors. To stabilize cycles of length $T$ the control system should consist of $l+1$ levels of mixing, the outside level is determined by the vector of the mixing coefficients of the $M_1$ dimension, the next level is determined by the mixing coefficients matrix of the $M_2\times M_1$ dimension, etc. Each level is determined by the tensor of the corresponding order  coefficients. All tensors must satisfy the folding conditions, in which sequential folding of the coefficients tensors from the first mixing level to $l=1$ level should give a vector of optimal coefficients $$\left(\begin{array}{c}
a_1\\
\ldots\\
a_N\\
\end{array}
\right).$$
It is clear that the optimal mixing coefficients are determined not always in one way. For example, one possible system of stabilization of the cycle length $2T$ can be the system
$$x_{n+1}=c_1\sum_{j=1}^{N}a_jf(f(x_{n-jT+T}))+c_2f(\sum_{j=1}^{N}a_jf(x_{n-jT+T}))+c_3f(f(\sum_{j=1}^{N}a_jx_{n-jT+T})),$$
where $c_1\ge 0,\,\,c_2\ge 0,\,\,c_3\ge 0,\,\,c_1+c_2+c_3=1$.
 
Note important differences between the stabilization method considered in this paper  and the most known methods. Control is applied at all points in time, and not just in a neighborhood of the desired cycle -- {\it the cycle to know in advance is not necessary}. Moreover, our control allows to stabilize at once ALL cycles of a given length with multipliers lying in the left half-plane, real or complex. To stabilize the specific cycle is sufficient for the sequence of initial points to lie in the basin of attraction of this cycle. Different initial sequence will generate sequences of solutions converging to the different cycles of a given length.
 
One of the possible applications of the proposed method is to check the existence of periodic orbits of a given nonlinear mapping, the unstable orbits can be detected by their stabilization.
 
\section{Optimal stabilization of the chaos}
 
Consider the mapping \cite{20} $$f(x)=(1+\sqrt2)\left(\frac{1}{2}-\left|x-\frac{1}{2}\right|\right)+x,$$
generating a dynamic system of "sudden occurrence of chaos" (SOC)
\begin{equation}
                x_{n+1}=f(x_n).
\end{equation}
In \cite{19} to stabilize (locate) the cycles of lengths from 1 to 6  the predictive control \cite{4} was used. Below  the application of the mixing  method for stabilize the 3-cycle (Fig. 1) , 7-cycles (Fig. 2), 9-cycles (Fig. 3), 12-cycles (Fig. 4), 15-cycles (Fig. 5) in the SOC system
is illustrated.\\
 
For the system (11) the invariant set is $\left[0,1+\frac{\sqrt2}{2}\right];$ the  equilibrium is point $z=1.$ \\
 
Let us find  3-cycle of system (11). To do that let us consider the control system
\begin{equation}
x_{n+1}=f\left(f\left(f\left(\sum_{j=1}^{N}a_jx_{n-jT+T}\right)\right)\right)
\end{equation}
with $T=1,\,\,N=8,\,\,\{a_1,\ldots,a_8\}=\{0.107,0.176,0.204,0.193,0.154,0.102,0.050,0.013\}$.
\begin{center}
               %\begin{picture}(13.5,5)
\includegraphics[width=14.5cm]{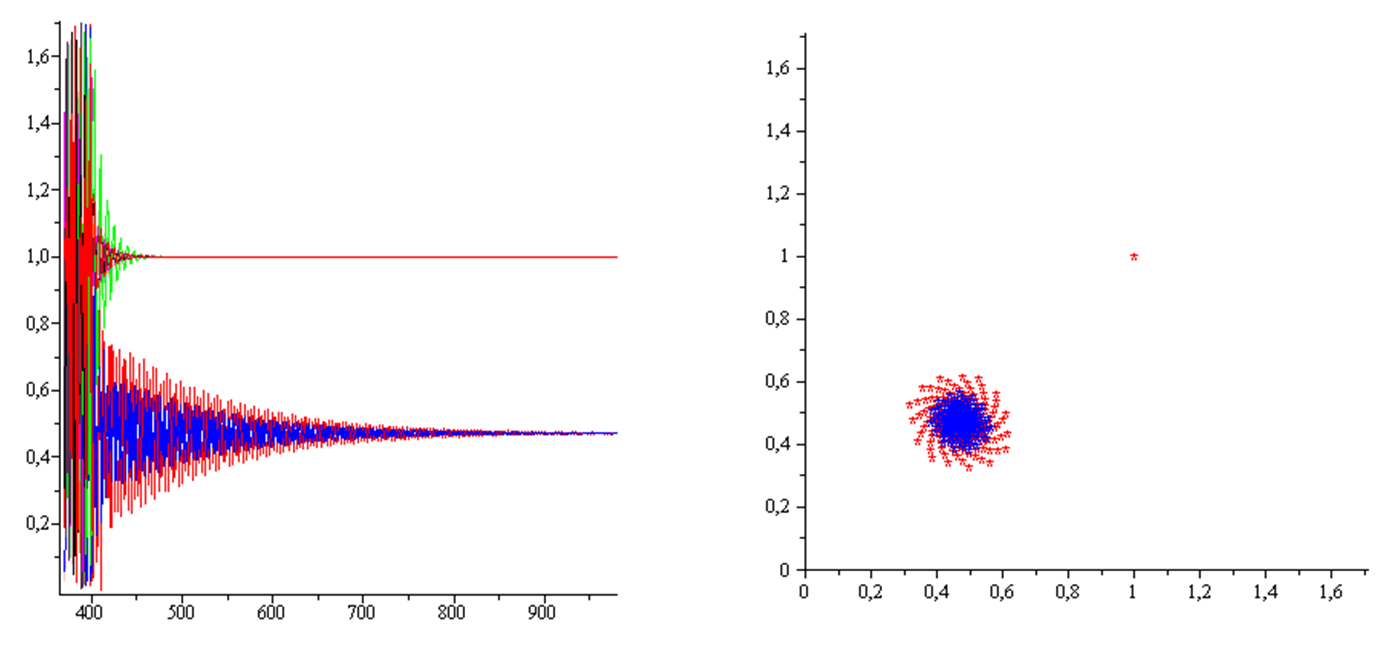}
               %\end{picture}

                Figure 1 - Dynamics of solutions of system (12) in $(n,x_n)$ and in $(x_n,x_{n+1})$ planes.
                \end{center}
 
We can see that system (12) has two equilibrium positions: $z_1=1$ and $z_2\approx 0.471.$ The first equilibrium position is the equilibrium position for the original system (11), but the second defines a 3-cycle system of the system (11): $\{0.471,1.609,0.138\}$.\\
 
Similarly, we can find the 7-cycle system (11). The control system
\begin{equation}
x_{n+1}=f\left(f\left(f\left(f\left(f\left(f\left(f\left(\sum_{j=1}^{N}a_jx_{n-jT+T}\right)\right)\right)\right)\right)\right)\right)
\end{equation}
uses $T=1,\;N=12$ and
$$
\{a_1\ldots,a_{12}\}=
\{0.054,0.095,0.124,0.138,0.139,0.130,0.111,0.087,0.061,0.037,0.017,0.004\}.$$
 
\begin{center}
               \includegraphics[width=14.5cm]{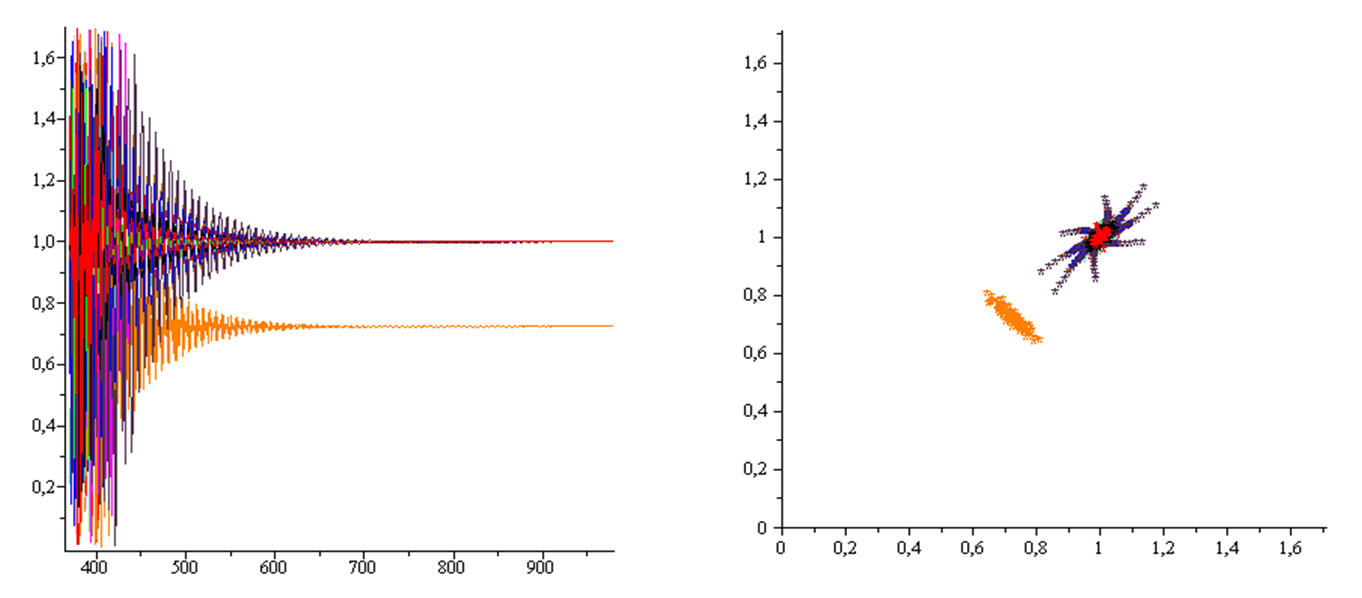}
 
                Figure 2 - Dynamics of solutions of system (13) in
               $(n,x_n)$ and $(x_{n-1},x_n)$ planes.
\end{center}
 
Position of equilibrium in the system (13) $z\approx0.723$ determines the desired 7-cycle system (11): $\{0.723,1.391,0.448,1.529,0.252,0.862,1.195\}$ .
 
Define the 9-cycle of SOC system. In the control system (12) suppose $T=3,$  $N=11,$ $$\{a_1,\ldots,a_{11}\}=\{0.276,0.168,0.127,0.102,0.084,0.070,0.058,0.046,0.036,0.024,0.009\}.$$
 
\begin{center}
               \includegraphics[width=14.5cm]{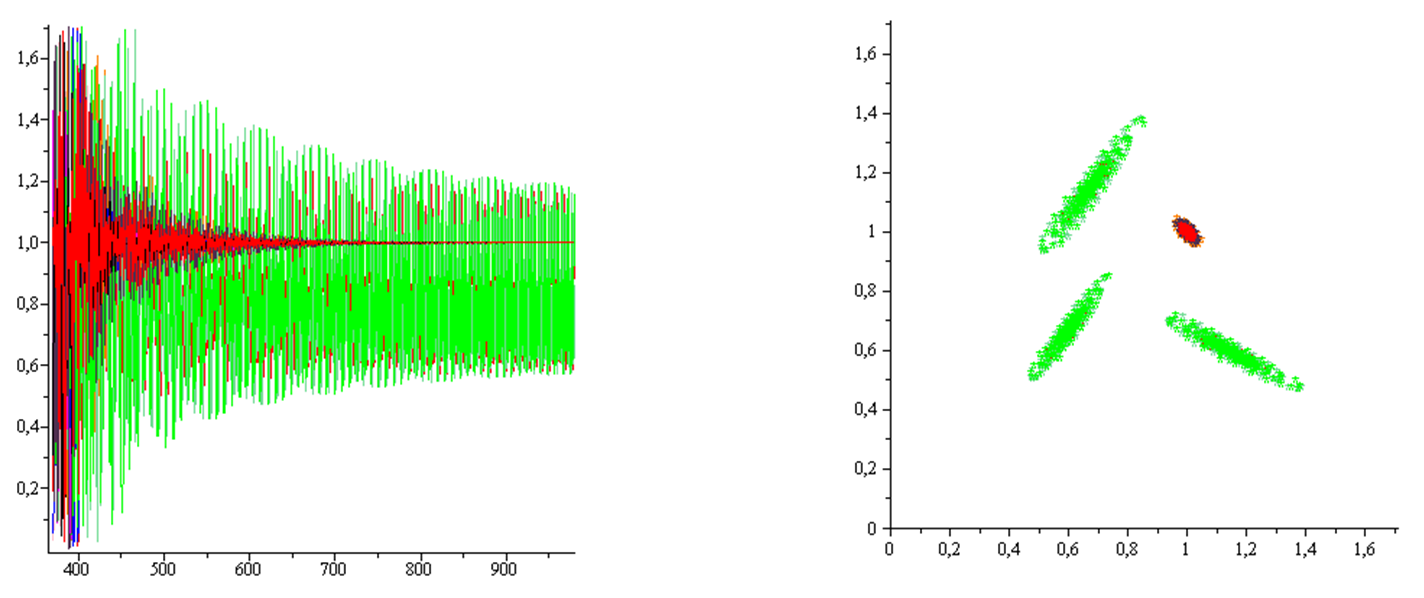}
 
                Figure 3 - Dynamics of solutions of system (12) with $T=3,\,\,N=11$ in planes $(n,x_n)$ and in $(x_{n-1},x_n)$.
\end{center}
 
In the control system (12) there is a locally stable equilibrium position, which is also the equilibrium position of the original system (11), and a locally stable 3-cycle $\{1.142,0.598,0.0667\}$. This 3-cycle determines the 9-cycle for system (11): $\{1.142,0.799,1.285,0.598,1.570,0.194,0.667,1.478,0.326\}.$
 
To determine  12-cycle of the system (11) let us applied the system
\begin{equation}
x_{n+1}=f\left(f\left(\sum_{j=1}^{N}a_jf(x_{n-jT+T})\right)\right)
\end{equation}
with $T=4,\,N=11$ and
$$\{a_1,\ldots,a_{11}\}=\{0.361,0.165,0.113,0.087,0.069,0.057,0.047,0.039,0.031,0.022,0.009\}.$$
 
\begin{center}
                \includegraphics[width=15cm]{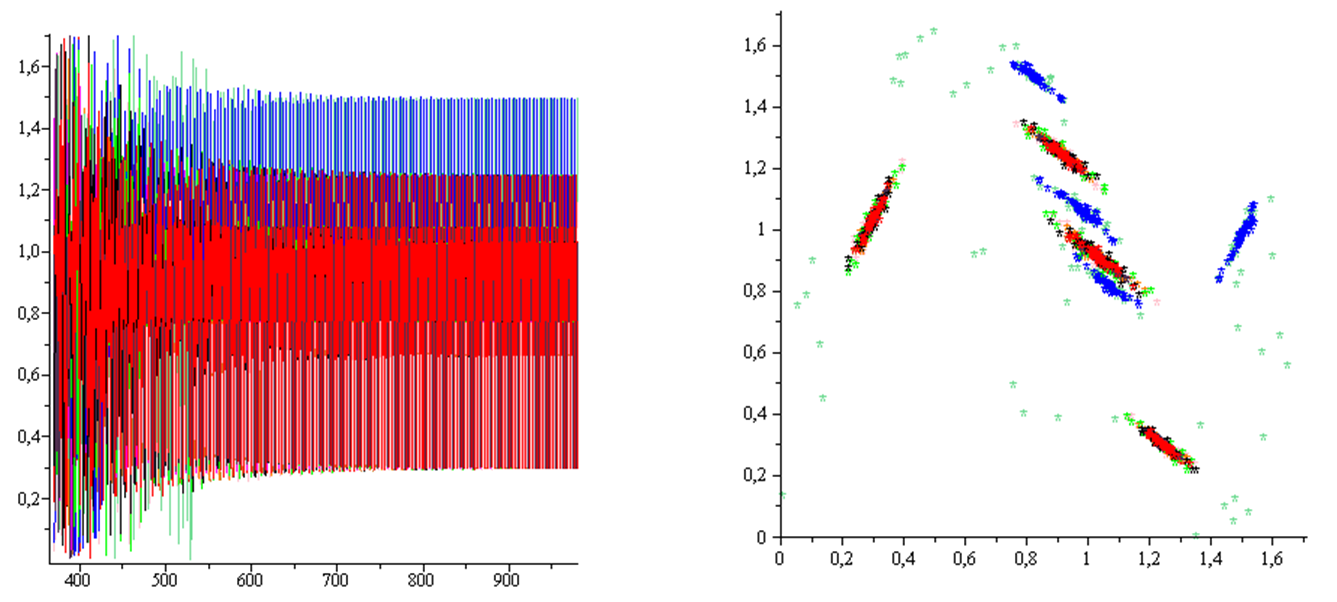}
 
                Figure 4 - Dynamics of solutions of system (14) in planes $(n,x_n)$ and $(x_{n-1},x_n)$.
\end{center}
 
In system (14) one can observe two locally stable 4-cycles. They are part of 12-cycle of the original system (11): $$\{1.031,0.956,1.062,0.912,1.124,0.824,1.248,0.649,1.497,0.297,1.016,0.978\}.$$
 
Generally speaking, with the increase of cycle length $T$, the number of cycles increases. Basins of attraction of some cycles can be so small that practically the cycles will be difficult to detect. Suppose, for example, one needs to find cycles of length 15. We choose several starting points and consider their dynamics in accordance with the equation
$$x_{n+1}=f(f(f(x_n))).$$
Take 400 iterations of this system, and in step 401  we start the mixing process (12). In this case let $T=5,\,\,N=26$, and the mixing coefficients will be chosen in accordance with the formulas of section 2. The dynamics of solutions of the system with different initial values is shown on Fig. 5. At some initial values of the trajectory will be attracted to the cycles of length 5: when $x_0=0.99$ we get cycle $\{0.989,1.025,0.929,1.198,0.0440\}$; when $x_0=0.64$ we get cycle $\{0.290,0.982,1.049,$ $0.860,1.96\}$; when $x_0=0.5$ and $x_0=0.74$ we get cycle $\{1.127,0.640,0.958,1.119,0.665\}$; when $x_0=0.21$ we get cycle $\{0.960,1.112,0.683,1.252,0.287\}$. These four 5-cycles generate four different 15-cycles of the original system (11). Namely,
$$\{0.989,1.509,0.280,1.025,0.963,1.052,0.929,1.100,0.859,1.198,0.719,1.397,0.440,1.501,0.291\}$$
$$\{0.290,0.991,1.013,0.982,1.025,0.965,1.049,0.930,1.099,0.860,1.198,0.720,1.396,0.440,1.502\}$$
$$\{1.127,0.820,1.254,0.640,1.509,0.281,0.958,1.059,0.916,1.119,0.832,1.237,0.665,1.474,0.330\}$$
$$\{0.960,1.056,0.921,1.112,0.842,1.224,0.683,1.448,0.367,1.252,0.644,1.504,0.287,0.981,1.027\}$$
 
\begin{center}
                \includegraphics[width=14.5cm]{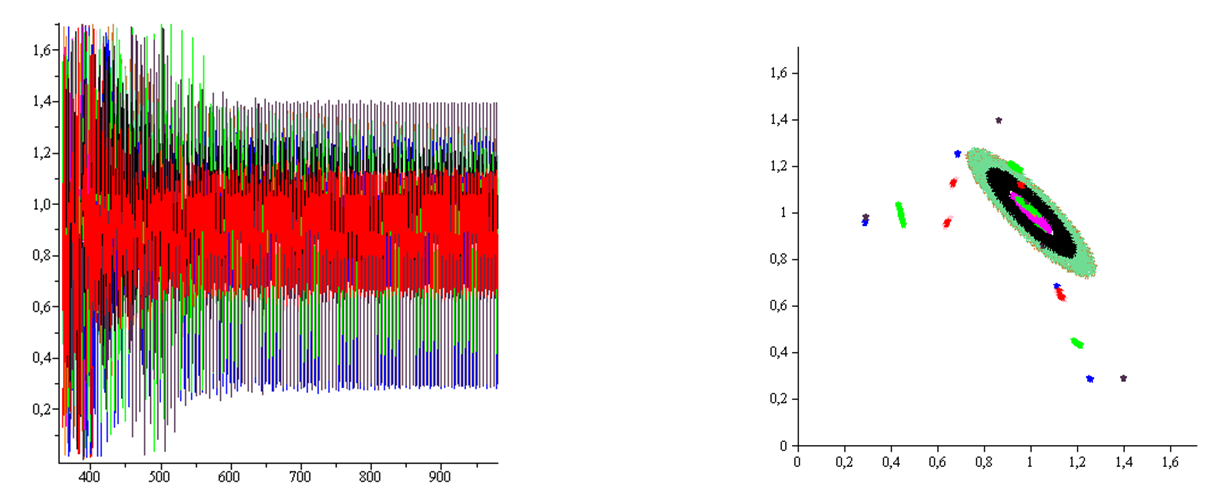}
 
                Figure 5 - Dynamics of solutions of system (12) in $(n,x_n)$ and in $(x_{n-1},x_n)$ planes with $T=5$ and $N=26.$
\end{center}
 
\section{Wolfram Code}
 
Stephen Wolfram, a founder of Wolfram Mathematica, conducted in the earlier stage of his career research in cellular automata. Wolfram code is a naming system often used for one-dimensional cellular automata rules, introduced by Wolfram in a 1983 paper \cite{21} and used in his book ``A New Kind of Science'' \cite{22}.  The rules defined an evolution that can be written in the form of dynamical system with internal mixing (see formula (2.2) in \cite{23})
$$
x_i(n+1)=f\left(  \sum_{j=-r}^r b_jx_{i+j}(n)\right),\; i=1,...,Q.
$$
So, the developed above theory can by applied to study Wolfram Code. \\
 
The above formula is an average in space coordinates with special rules at end points. Those extensions can be performed in many ways. For example, by cyclic advances, by mirror symmetry, etc.
 
Now, let us apply time averaging developed in section 2. It can be done for example with internal mixing
$$
x_i(n+1)=f\left(  \sum_{k=1}^N\sum_{j=-r}^r a_kb_jx_{i+j}(n-kT+T)\right),
  $$
or external mixing
$$
x_i(n+1)=\sum_{k=1}^N a_k f\left(  \sum_{j=-r}^r b_jx_{i+j}(n-kT+T)\right).
  $$
These mixing could detect periodical or stationary regimes for dynamics of cellular automata.

\section{Conclusion}
 
This paper considers the important problem of chaos control by local stabilizing of a priori unknown unstable periodic orbits of discrete systems with chaotic dynamics. The approach for this stabilization problem uses control that involves the mixing of coordinates of the prehistory or functions of these coordinates. This  approach generalizes the method of stabilization using nonlinear delayed feedback control (DFC) proposed in \cite{15, 16}. It retains all the advantages of the DFC method and provides additional opportunity for the choosing of  mixing parameters, which is extremely important for nonlinear systems, allowing one to expand the basins of attraction of stabilized periodic orbits. Note, however, that the use of different mixing levels does not allow one to reduce the length of the background needed to stabilize the cycle. But one can increase the rate of convergence to this cycle. In addition, using inside mixing (the mixing of the first level) significantly reduces the amount of computation, because there is no need to compute a value of a function that specifies the dynamic system, greater than number of times than in the system without mixing. We also hope that using first level mixing will facilitate the physical implementation of the proposed scheme of controlling chaos in nonlinear discrete systems.
 
\section{Acknowledgment}
 
The authors would like to thank Paul Hagelstein and Emil Iacob  for interesting discussions and  for the help in preparation of manuscript.

\bigskip

D. Dmitrishin and  I.M. Skrinnik, Odessa National Polytechnic University,  1 Shevchenko Ave, Odessa 65044, Ukraine. E-mail: dmitrishin@opi.ua\\

A. Stokolos, Georgia Southern University, Statesboro, GA 3040, USA. E-mail:\\ astokolos@georgiasouthern.edu

\end{document}